\documentstyle[aps]{revtex}
\draft

\begin{document}

\title{The Energy of a Trapped Interacting Bose Gas}
\author{Hualin Shi and Wei-Mou Zheng}
\address{Institute of Theoretical Physics, Academia Sinica, Beijing
100080,China}
\maketitle

\begin{abstract}
A Bose gas in an external potential is studied by means of the
semi-classical approximation. Analytical results are derived for the
energy of an interacting Bose gas in a generic power-law trapping
potential. An expression for the chemical potential below the critical
temperature is also obtained. The theoretical results are in qualitative
agreement with a recent energy measurement.
\end{abstract}


The condensation of an ideal Bose-Einstein gas is one of the most striking
consequences of quantum statistics \cite{huang87}. When Bose-Einstein
condensation (BEC) occurs at a sufficiently low temperature, the zero
momentum state can become macroscopically occupied. For many years, it
was considered hopeless to experimentally observe BEC in an
atomic gas with weak interactions. With the development of techniques to
trap and cool atoms, BEC was recently observed directly in dilute atomic
vapors\cite{and95,bra95,dav95}. The new experimental achievements have
stimulated great interest in the theoretical study of inhomogeneous Bose
gases.

The thermodynamic properties of trapped atomic Bose gases undergoing BEC can
be altered by the spatially varying trapping potential. The interaction
between atoms may have a significant effect on the thermodynamic properties.
There have been several investigations analyzing the dependence of the
critical temperature on the trapping potential and weak interaction in the
Bose gas\cite{bag87,gio96,shi96b}. The thermodynamic properties of Bose
gases in an external potential have also been discussed in
Refs.~\cite{bag87,shi96c}.

In a recent experiment\cite{ens96}, after turning off the trapping field,
the kinetic energy of a sufficiently expanded atom cloud was measured.
The experimental data is in a good agreement with the theory of a trapped
ideal Bose gas when the temperature is above critical temperature. However,
a discrepancy with the theory of ideal Bose gases is found below the
critical temperature. To explain this discrepancy it is important to know
how the mutual interaction affects the energy. Here we shall derive some
analytical expressions for the energy of a trapped non-interacting and
interacting Bose gas under the semi-classical approximation.

First, let us calculate the kinetic and potential energies of an ideal
Maxwell-Boltzmann gas and of an ideal Bose-Einstein gas in an external
potential. For an ideal Maxwell-Boltzmann gas, according to the
Maxwell-Boltzmann distribution, the local atom number density is given by
\begin{equation}
\label{stat-mb}
n_{\hbox{\tiny MB}}({\bf r,p},T)= \exp [-\beta
(p^2/2m+V({\bf r})-\mu )] ,
\end{equation}
where $\beta=kT$ and $\mu$ is the chemical potential which comes from the
normalization. Integration of the number density
(\ref{stat-mb}) over the whole phase space gives the total number of atoms
\begin{equation}
N = \frac 1{h^3}\int n_{\hbox{\tiny MB}}({\bf r,p},T)d{\bf p}d{\bf r}
= \frac{z}{\lambda ^3}\int \exp [-\beta V({\bf r})]d{\bf r} ,
\label{n-mb}
\end{equation}
where
\begin{equation}
\lambda= \left(\frac{2 \pi \hbar^2}{m kT} \right)^{1/2},
\qquad z = e^{\beta \mu} .
\end{equation}
For the generic power-law potential discussed in Ref.~\cite{bag87}
\begin{equation}
\label{pot}V({\bf r})=\epsilon _1\left| \frac x{L_1}\right| ^p+\epsilon
_2\left| \frac y{L_2}\right| ^l+\epsilon _3\left| \frac z{L_3}\right| ^q,
\end{equation}
where $L_1$, $L_2$ and $L_3$ are the linear sizes of the volume, and
$\varepsilon_1$, $\varepsilon_2$, $\varepsilon_3$, $p$, $l$ and $q$ the
parameters, Eq.~(\ref{n-mb}) yields
\begin{equation}
\label{nt}
N=z\lambda ^{-3} \chi I(p,l,q)\beta^{-\eta +1/2} ,
\end{equation}
with
\begin{equation}
\chi = \frac{L_1 L_2 L_3}{\epsilon_1^{1/p}\epsilon_2^{1/l}\epsilon_3^{1/q}},
\qquad
I(p,l,q) = \frac{8}{plq}\; \Gamma (1/p)\;\Gamma (1/l)\;\Gamma (1/q), \qquad
\eta =\frac{1}{p}+\frac{1}{l}+\frac{1}{q} + \frac{1}{2} .
\end{equation}
In the derivation we have used the formula for the gamma function
\begin{equation}
\label{gamma}
\Gamma (z)=\int_0^\infty t^{z-1}e^{-t}\;dt.
\end{equation}
Relation (\ref{nt}) determines the chemical potential $\mu$ from a given
total number $N$ of atoms. The total kinetic energy can be obtained as
\begin{equation}
K_{\hbox{\tiny MB}} =\int \frac{p^2}{2m}\;dN({\bf r,p},T)
=\frac 32kT\frac{z}{\lambda ^3}\int \exp [-\beta V({\bf r})] \; d{\bf r}
=\frac 3 2 NkT,
\label{t-mb}
\end{equation}
which is just a result of the energy equipartition principle.
Similarly, the total potential energy is
\begin{equation}
V_{\hbox{\tiny MB}} = \int V({\bf r})\;dN({\bf r,p},T) =
\left( \frac 1p+\frac 1l+\frac 1q\right)
\frac{z \chi I(p,l,q)}{\lambda^3 \beta^{\eta+1/2}} .
\end{equation}
Thus, the total energy of the trapped ideal Maxwell-Boltzmann gas is
\begin{equation}
E_{\hbox{\tiny MB}}^{t}=\left( \eta +1\right) NkT .
\end{equation}

We next consider an ideal Bose-Einstein gas in the generic power law
potential. Under the semi-classcial approximation\cite{chou96,bag87,gio96}
the local number density is given by the Bose-Einstein distribution
function
\begin{equation}
\label{stat-be}
n_{\hbox{\tiny BE}}({\bf r,p},T)=
\{\exp [\beta (p^2/2m+V({\bf r})-\mu )]-1\}^{-1},
\end{equation}
from which the reduced spatial distribution of atoms is \cite{bag87,chou96}
\begin{equation}
\rho({\bf r}) = \lambda^{-3} g_{3/2}[\exp(-\beta (V(\bf{r})-\mu))],
\label{den-be}
\end{equation}
where $g_\nu(x)=\sum_{j=1} x^j/j^{\nu}$.

The critical temperature for an ideal Bose-Einstein gas in the generic
power law potential to undergo BEC has been found in
Refs.~\cite{bag87,shi96b} to be
\begin{equation}
T_c= \frac 1 k\;\left(\frac{N\;(2\pi \hbar ^2)^{3/2}}{\;m^{3/2}\;
\zeta (\eta +1)\; \chi\; I(p,l,q) }\right)^{1/(\eta+1)},
\label{i-t}
\end{equation}
where $\zeta (\nu )=g_\nu (1)$ is the Riemann zeta function.

The cases of the temperature below and above $T_c$ should be treated
separately. For $T>T_c$, integrating the number density function (\ref
{stat-be}) over the phase space, we obtain the total number of atoms
\cite{shi96c}
\begin{equation}
\label{ben}
N =h^{-3} \int n_{\hbox{\tiny BE}}({\bf r,p},T) d{\bf r}d{\bf p}
=\frac{\chi I(p,l,q) }{\lambda ^3\beta ^{\eta -1/2}} g_{\eta +1}(z).
\end{equation}
In the derivation we have used the definition of the function
$g_{\nu}$. Similarly, we have the total kinetic energy
\begin{equation}
K_{\hbox{\tiny BE}} =h^{-3}\int
\frac{p^2}{2m} n_{\hbox{\tiny BE}}({\bf r,p},T) d{\bf p}d{\bf r}
 =\frac 32 \frac{\chi I(p,l,q)}{\lambda ^3\beta ^{\eta +1/2}}\;g_{\eta
 +2}(z),
\label{t-be-i}
\end{equation}
and the total potential energy
\begin{equation}
V_{\hbox{\tiny BE}}  = \int V({\bf r})\rho ({\bf r}) d{\bf r}
 = \frac{\chi I(p,l,q)}
{\lambda ^3\beta ^{\eta +1/2}}\left( \frac 1p+\frac
1l+\frac 1q\right) g_{\eta +2}(z) .
\label{v-be-i}
\end{equation}
The total energy of the trapped ideal Bose gas is then\cite{pin96}
\begin{equation}
E_{\hbox{\tiny BE}}^{t}=\frac{g_{\eta +2}(z)}{g_{\eta +1}(z)}
\left( \eta +1\right) NkT
\label{tot-be-i}
\end{equation}
Comparing with the trapped ideal Maxwell-Boltzmann gas, we see that
the only difference in the two total energy expressions is the factor
$g_{\eta +2}(z)/g_{\eta +1}(z)$ for the Bose gas, which breaks the linear
dependence of the total energy on the temperature.

When the temperature is below $T_c$, the contribution to the energy of the
system comes only from the normal component of the Bose gas. From
Eqs.~(\ref{ben}), (\ref{t-be-i}), (\ref{v-be-i}) and (\ref{tot-be-i}) the
kinetic, potential and total energy of the system can be written as
\begin{eqnarray}
K_{\hbox{\tiny BE}} &=& \frac 32\frac{g_{\eta +2}(1)}{g_{\eta +1}(1)}\left(
\frac T{T_c}\right) ^{\eta +1}NkT ,
\label{t-bl-i}\\
V_{\hbox{\tiny BE}} &=& \left( \frac 1p+\frac 1l+\frac 1q\right)
\frac{g_{\eta +2}(1)}{g_{\eta +1}(1)}\left( \frac T{T_c}\right) ^{\eta +1}
NkT,
\label{v-bl-i}\\
E^{t}_{\hbox{\tiny BE}} &=& (\eta +1)\frac{g_{\eta +2}(1)}
{g_{\eta +1}(1)}\left( \frac T{T_c}\right)^{\eta +1} NkT .
\end{eqnarray}
The energy of the system will go to zero when the temperature approaches zero.

So far we have not considered the role of the mutual interaction of Bose
atoms. For a trapped interacting Bose gas, we divide the total energy into
three parts: the kinetic energy $K_{int}$, potential energy $V_{int}$ and
interaction energy $U_{int}$, i.e.
\begin{equation}
\label{en-tot}E^{t}_{int}=K_{int}+V_{int}+U_{int} ,
\end{equation}

In the local density approximation or semi-classical approximation, the
local number density function of a trapped interacting Bose gas is given
by \cite{gio96,chou96}
\begin{equation}
\label{stat-be-i}
n({\bf r,p},T)=\{\exp [\beta (p^2/2m+V({\bf r})+
4a\lambda ^2\rho ({\bf r})-\mu )]-1\}^{-1}.
\end{equation}

In similarity to the non-interacting Bose gas, the cases of the temperature
above and below $\tilde T_c$ need to be dealt with separately, where $\tilde
T_c$ is the critical temperature of the trapped interacting Bose gas
\cite{shi96b}. At a temperature above $\tilde T_c$, from (\ref{stat-be-i})
the total number of atoms is \cite{shi96c}
\begin{equation}
N =\frac 1{h^3}\int n({\bf r,p},T)d{\bf p}d{\bf r}
=\frac{\chi I(p,l,q) }{\lambda ^3\beta ^{\eta -1/2}} \left [
g_{\eta +1}(z) -\frac{2 a}{\lambda} F_{3/2,3/2,\eta -1}(z)
\right ] ,
\end{equation}
where
\begin{equation}
F_{\delta ,\nu ,\eta }(x) =\sum_{i,j}^\infty \frac{x^{i+j}} {i^\delta j^\nu
(i+j)^{\eta -1/2}} .
\end{equation}
In the derivation we have expanded the exponential density function with
respect to the small parameter $a/\lambda$, and kept only terms up to the
lowest order in $a/\lambda$ \cite{shi96b,shi96c}. Up to this first order
correction, Eq.~(\ref{den-be}) can be used. By the same procedure, the
total kinetic and potential energy are obtained as
\begin{eqnarray}
K_{int} &=&\frac 1{h^3}\int \frac{p^2}{2m}n({\bf r,p},T)d{\bf p}d{\bf r}
=\frac 32 \frac{\chi I(p,l,q) }{\lambda ^3\beta ^{\eta +1/2}} \left [
g_{\eta +2}(z) -\frac{4 a}{\lambda} F_{3/2,3/2,\eta}(z)
\right ] ,\\
V_{int}
&=& \frac{1}{h^3} \int V({\bf r})n({\bf r,p},T)d{\bf p}d{\bf r}
=\left(\frac{1}{p}+\frac{1}{l}+\frac{1}{q} \right) \frac{\chi I(p,l,q) }
{\lambda ^3\beta ^{\eta +1/2}}\left [ g_{\eta +2}(z)
-\frac{4 a}{\lambda} F_{1/2,3/2,\eta+1}(z) \right ] .
\end{eqnarray}
Similarly, with $V({\bf r})$ being replaced by $2 a\lambda ^2kT\rho ({\bf r})$,
the interaction energy is calculated as
\begin{equation}
U_{int} =\int 2 a\lambda ^2kT\rho ^2({\bf r})d{\bf r}
 =\frac{2 a \chi I(p,l,q) }{\lambda ^4\beta ^{\eta +1/2}}
F_{3/2,3/2,\eta }(z) .
\end{equation}
From the above three expressions for $K_{int}$, $V_{int}$ and $U_{int}$,
we see that the correction to the energy of the trapped ideal Bose gas due
to the mutual interaction of atoms is proportional to $a/\lambda$, hence is
very small for a weak interaction. This explains why the ideal Bose gas
result agrees well with the experimental data above $\tilde T_c$.

When the temperature is below $\tilde T_c$, a region where a great number
of atoms are in the BEC state forms. In the local density approximation, a
relation between density and potential inside the region reads \cite{chou96}
\begin{equation}
V({\bf r})+4\pi a\rho ({\bf r})\hbar ^2/m=V({\bf r}_0)+4\pi a\rho _0\hbar
^2/m, \qquad r<r_0 ,
\end{equation}
where $r_0$ corresponds to the edge of the region. At $r = r_0^-$, we
have $\rho({r_0^-}) =\rho_0=\lambda^{-3} g_{3/2}(1)$, which is the critical
density of the gaseous phase for the untrapped homogeneous ideal Bose gas
at temperature $T$. From the above relation, we find the density of the
condensed atoms as the difference
of $\rho({\bf r})$ and $\rho_0$
\begin{equation}
\rho _s({\bf r})=\rho ({\bf r})-\rho_0 =\frac{m}{4 \pi a \hbar^2}
[V({\bf r_0})-V({\bf r})],\qquad r<r_0 .
\label{den-bec}
\end{equation}
In order to obtain $\rho_s$, the potential $V(r_0)$ at the edge of the
condensed region needs to be determined. Integrating $\rho_s$ over the
whole volume, we have the total number of atoms in the condensed state
\begin{equation}
N_s = \frac{m \chi }{4 \pi a \hbar^2} \frac{I(p,l,q)}{\Gamma (\eta +3/2)}
V^{\eta+1/2}(r_0) ,
\label{ns-c}
\end{equation}
where the derivation involves the integral
\begin{equation}
\int_{|x|^p+|y|^l+|z|^q\leq 1} (1-|x|^p-|y|^l-|z|^q)\;dxdydz
=\frac{I(p,l,q)}{\Gamma (\eta +3/2)}.
\end{equation}
At the weak interaction limit we may approximate $N_s$ and $\tilde T_c$ by
those for the trapped ideal Bose gas \cite{bag87,shi96b}, i.e.
\begin{equation}
N_s \approx N \left [ 1-\left(\frac{T}{T_c} \right)^{\eta+1} \right ]
,\qquad T<T_c,
\label{ns-t}
\end{equation}
where we have written $\tilde T_c$ as $T_c$. From Eqs.~(\ref{ns-c}) and
(\ref{ns-t}) we find
\begin{equation}
\label{v0}
V(r_0) = \left \{ \frac{4 \pi a \hbar^2\Gamma (\eta +3/2)}{m \chi I(p,l,q)}
\left[1-\left(\frac{T}{T_c} \right)^{\eta+1} \right ]N\right\}^{2/(2\eta+1)}.
\end{equation}
For the condensation region at $T<T_c$ there exists another relation
\cite{chou96}
\begin{equation}
kT \ln z - V({\bf r}) = \frac{4 \pi a \hbar^2}{m} [\rho({\bf r})+
\rho_0 ] , \qquad  (r<r_0) .
\end{equation}
Setting $r=r_0^-$ in this formula, from Eq.~(\ref{v0}) we find the
chemical potential as a descending function of temperature to be
\begin{equation}
\mu = \frac{8 \pi a \hbar^2}{m} \rho_0
+\left \{ \frac{4 \pi a \hbar^2\Gamma (\eta +3/2)}{m \chi I(p,l,q)}
\left[1-\left(\frac{T}{T_c} \right)^{\eta+1} \right ]
N \right \}^{2/(2\eta+1)},\qquad (T<T_c) ,
\label{ch-bl}
\end{equation}
which, for a cylindrically symmetric harmonic trapping potential,
reduces to
\begin{equation}
\mu = \frac{8\pi a\rho _0\hbar ^2}m+\frac{\hbar \omega_0 }2\left\{
\frac{15aN}{a_0}\left[ 1-\left( \frac T{T_c}\right) ^3\right] \right\} ^{2/5}
\buildrel {T\to 0}\over\longrightarrow
\frac{\hbar \omega_0 }2\left\{ \frac{15aN}{a_0}\right\} ^{2/5},
\end{equation}
where $a_0$ is the geometric mean characteristic length of the trap,
i.e. $a_0^2=\hbar /(m\omega_0)$ with $\omega_0^3=\omega_{\perp}^2 \omega_z$.
This is consistent with the Ginzburg-Pitaevskii-Gross mean field theory
\cite{bay96}.

According to Refs.\cite{chou96,huang57}, the density of the mutual interaction
energy in the condensation region is
\begin{eqnarray}
u_{int} &=& 2 a \lambda^2 kT \rho^2({\bf r})
-a \lambda^2 kT \rho_s^2({\bf r}) \nonumber \\
&=& a \lambda^2 kT [2 \rho_0^2({\bf r}) +4 \rho_0 \rho_s({\bf r})
+\rho_s^2({\bf r}) ],\qquad (r<r_0),
\end{eqnarray}
which, after integrating over the condensation region ($r<r_0$), gives
\begin{equation}
U_{int} = \chi I(p,l,q)V^{\eta +1/2}(r_0)\; \left[ \frac{4 \pi a \hbar^2
\rho_0^2}{m\Gamma (\eta+1/2) V(r_0)} +\frac{2 \rho_0}{\Gamma (\eta+3/2)}
+\frac{m V(r_0)}{4\pi a \hbar^2 \Gamma (\eta+5/2)}\right].
\label{en-int}
\end{equation}
The first two terms depending on $\rho_0$ vanish when the temperature 
approaches zero. So, at low temperature the main contribution to $U_{int}$ 
comes from the third term corresponding to the interaction within the 
condensed atoms. We may then neglect the first and second terms 
of (\ref{en-int}) to write
\begin{equation}
U_{int} \approx \frac{1}{\Gamma (\eta+5/2)}
\left(\frac{4 \pi a \hbar^2}{m\chi I(p,l,q)}\right)^{2/(2\eta +1)}
\left\{\left[1-\left(\frac{T}{T_c} \right)^{\eta+1}
\right ]\Gamma (\eta+3/2)N \right \}^{(2 \eta +3)/(2\eta+1)}.
\label{int-bec}
\end{equation}
To an approximation, we may neglect any correction to the energy from
the normal component, and express the energy of the trapped interacting Bose
gas as that of the trapped ideal Bose gas plus the only correction from the
mutual interaction within the condensed component.

In the experiment the trapping potential is \cite{ens96}
\begin{equation}
V({\bf r})=\frac 12m\omega_{\perp}^2r_{\perp}^2
+ \frac{1}{2} \omega_z^2 r_z^2,
\end{equation}
which is harmonic. In the ideal Bose gas approximation, at a temperature
$T$ above $T_c$, Eqs.~(\ref{t-be-i}) and (\ref{v-be-i}) reduce to
\begin{equation}
K = \frac{3}{2} \frac{g_4(z)}{g_3(z)}NkT,
\qquad
V = \frac{3}{2} \frac{g_4(z)}{g_3(z)}NkT,
\end{equation}
where the fugacity $z$ is related to the total number of atoms as
\begin{equation}
N = \frac{g_3(z)}{\hbar^3 \omega_{\perp}^2 \omega_z \beta^3} ,
\end{equation}
which yields $g_3(z)=(T_c/T)^3 g_3(1)$. The energy of the atom interaction
is negligible at $T>T_c$.

When the temperature is below $T_c$, from Eqs.~(\ref{t-bl-i}) and
(\ref{v-bl-i}), the approximate kinetic and potential energy are
\begin{equation}
K^-= \frac{3}{2} \frac{\zeta(4)}{\zeta(3)}
\left(\frac{T}{T_c} \right)^3 NkT ,\qquad V^- = K^- .
\end{equation}
The contribution to the energy from the interaction within the condensation
component is now significant. From Eqs.~(\ref{i-t}) and (\ref{int-bec}),
this interaction energy is
\begin{equation}
U^-=\frac{1}{7}
\left( \frac{15a}{a_0}\right) ^{2/5}N^{1/15} (\zeta(3))^{1/3} \left[
1-\left( \frac T{T_c}\right) ^3\right] ^{7/5} N k T_c .
\end{equation}

In the experiment the trapped atoms are $^{87}$Rb \cite{ens96}. The
$s$-wave scattering length of $^{87}$Rb is $a \approx 100$ in the unit of
the Bohr radius $a_B$. The trapping harmonic potential is axially symmetric
with the frequency ratio of axial to radial being $\sqrt{8}$. The final
stage of the evaporative cooling is performed at $\nu_z = 373$ Hz. The
trapping field is then turned off to initiate the expansion of the atom
cloud non-adiabatically. The kinetic energy $E_{\hbox{\tiny exp}}$ of atoms
is measured after a sufficient expansion which ensures the interaction
energy to convert to purely kinetic energy. Therefore, we may estimate
$E_{\hbox{\tiny exp}}$ as the sum of $K^-$ and $U^-$, i.e.
\begin{equation}
\frac{E_{\hbox{\tiny exp}}}{N kT_c} = \frac{3}{2}\frac{g_4(z)}
{g_3(z)} \left( \frac{T}{T_c} \right),  \hbox{\quad for\quad} T>T_c,
\end{equation}
and
\begin{equation}
\frac{E}{N kT_c} = \frac{3}{2}\frac{g_4(1)} {g_3(1)}
\left( \frac{T}{T_c} \right)^4 + 0.449 \left(\frac{a}{a_0} \right)^{2/5}
N^{1/15} \left[ 1-\left(\frac{T}{T_c}\right)^3 \right]^{7/5}, \hbox{\quad
for\quad} T<T_c .
\end{equation}
Assume that the total number of atoms in the trap is 40000. The scaled
energy per particle, $E/NkT_c$, is plotted versus the scaled temperature,
$T/T_c$, in Fig.~1, where the solid, dashed and dotted lines correspond to
the ideal Bose gas, ideal Maxwell-Boltzmann and interacting Bose gas,
respectively. When the temperature is rather high, they are very
close. However, at low temperature the differences among them are prominent.
The theoretical result of the interacting Bose gas is in qualitative
agreement with the experimental data. We have considered only the contribution 
from the mutual interaction of the condensed atoms. If the finite size 
effect\cite{gro95,ket96} and all other interaction terms are included, a 
quantitative comparison with experimental data can be made.

\acknowledgments
{The authors thank Hao Bai-lin for his encouragement and useful discussions.
This work was supported in part by the National Natural Science Foundation
of China.}

\bigskip
\leftline{\bf Figure Captions}
\medskip
\noindent Fig.~1 Plot of the scaled energy per particle $E/NkT_c$ of
trapped gas vs.\ scaled temperature $T/T_c$. The solid, dashed and
dotted lines correspond to an ideal Bose gas, ideal Maxwell-Boltzmann gas
and interacting Bose gas, respectively.

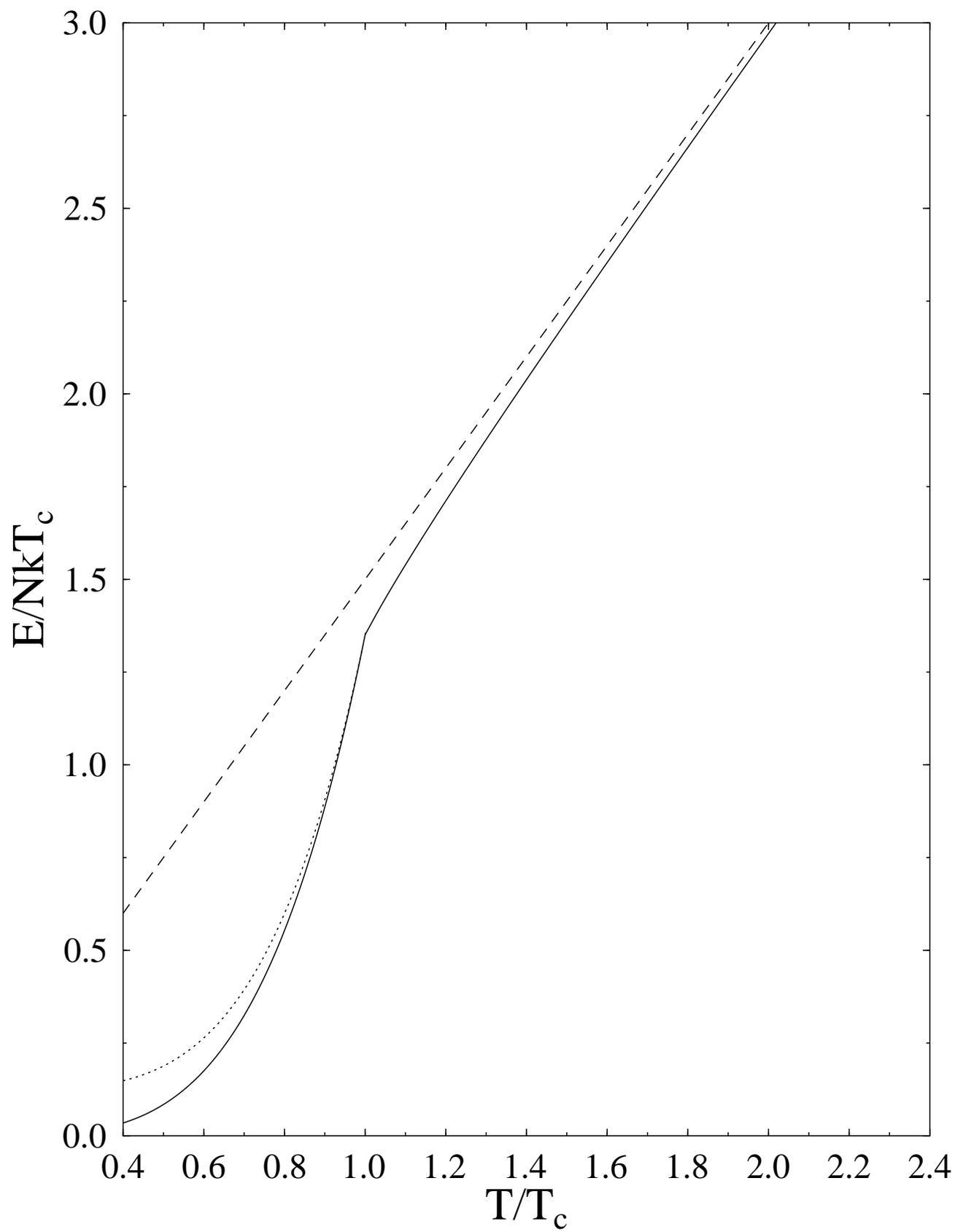
\end{document}